\documentclass{aims}
\usepackage{amsmath}
  \usepackage{paralist}
  \usepackage{graphics} %% add this and next lines if pictures should be in esp format
  \usepackage{epsfig} %For pictures: screened artwork should be set up with an 85 or 100 line screen
\usepackage{graphicx}  \usepackage{epstopdf}%This is to transfer .eps figure to .pdf figure; please compile your paper using PDFLeTex or PDFTeXify.
\usepackage{setspace}
\usepackage{verbatim}
\usepackage{amsfonts}
\usepackage{amssymb}

 \usepackage[colorlinks=true]{hyperref}
\hypersetup{urlcolor=blue, citecolor=red}

\def\currentvolume{X}
 \def\currentissue{X}
  \def\currentyear{200X}
   \def\currentmonth{XX}
    \def\ppages{X--XX}
     \def\DOI{10.3934/xx.xx.xx.xx}

\theoremstyle{definition}

\newcommand{\be}{\begin{equation}}   
\newcommand{\ee}{\end{equation}}

\newcommand{\bea}{\begin{eqnarray}}  
\newcommand{\eea}{\end{eqnarray}}
\newcommand{\bean}{\begin{eqnarray*}}
\newcommand{\eean}{\end{eqnarray*}}
\newcommand{\bit}{\begin{itemize}}   
\newcommand{\eit}{\end{itemize}}
\newcommand{\coefunit}{{\rm mm}^{-1}}
\def\mua{\mu_{{\rm a}}}
\def\mus{\mu_{{\rm s}}}

\newcommand{\R}{\mathbb{R}}

\def\iint{\int\kern-5pt\int\kern2pt}
\def\iiiint{\int\kern-5pt\int\kern-5pt\int\kern-5pt\int\kern2pt}

\def\mtrx#1#2{
  \left(
    \begin{array}{#1}
      #2
    \end{array}
  \right)}
\newcommand{\dontshow}[1]{}
\def\mua{\mu_{{\rm a}}}
\def\mus{\mu_{{\rm s}}'}
\newcommand{\musbar}{\mu_{{\rm s,\ast}}'}
\newcommand{\muabar}{\mu_{{\rm a,\ast}}}

\newcommand{\haem}{h_{{\rm aem}}}  
\newcommand{\hcem}{h_{{\rm cem}}}  
\newcommand{\hreff}{h_{{\rm ref}}}

\newcommand{\rmd}{{\rm d}}

\newcommand{\domain}{\Omega}

\newcommand{\mm}{{\rm mm}}

\newcommand{\nmeas}{m}

\newcommand{\ape}{\varepsilon}

\newcommand{\yfo}{y^{\rm f}_{\rm obs}}
\newcommand{\yeo}{y^{\rm e}_{\rm obs}}
\newcommand{\yec}{y^{\rm e}_{\rm calc}}
\newcommand{\transp}{^{\rm T}}

\newcommand{\submua}[1]{ \mu_{{\mathrm a},{#1}}}        
\newcommand{\submus}[1]{ \mu_{{\mathrm s},{#1}}'}  

\title[Approximate marginalization of ($\mu_a$, $\mu_s'$) in fDOT] %Use the shortened version of the full title
      {Approximate marginalization of absorption and scattering in fluorescence diffuse optical tomography}

\author[Mozumder Tarvainen Kaipio Arridge D'Andrea and Kolehmainen]{}

\subjclass{Primary: 74J25, 92C55; Secondary: 65R32.}
 \keywords{Image reconstruction techniques,  Tomography, Inverse problems, Bayesian methods, Fluorescence diffuse optical tomography.}

 \email{meghdoot.mozumder@uef.fi}
 \email{tanja.tarvainen@uef.fi}
 \email{jari@math.auckland.ac.nz}
 \email{cosimo.dandrea@polimi.it}
 \email{ville.kolehmainen@uef.fi}

\begin{document}
\maketitle

% Enter the first author's name and address:
\centerline{\scshape Meghdoot Mozumder and Tanja Tarvainen}
\medskip
{\footnotesize
% please put the address of the first author
 \centerline{Department of Applied Physics, University of Eastern Finland, P.O. Box 1627, 70211 Kuopio, Finland}
} % Do not forget to end the {\footnotesize by the sign }

\medskip

\centerline{\scshape Simon Arridge}
\medskip
{\footnotesize
 % please put the address of the second  and third author
 \centerline{Department of Computer Science, University College London, Gower Street, London WC1E 6BT, UK}
}

\medskip

\centerline{\scshape Jari P. Kaipio}
\medskip
{\footnotesize
% please put the address of the first author
 \centerline{Department of Mathematics, University of Auckland, Private Bag 92019, Auckland Mail Centre, Auckland 1142, New Zealand}
} % Do not forget to end the {\footnotesize by the sign }

\medskip

\centerline{\scshape Cosimo D' Andrea}
\medskip
{\footnotesize
% please put the address of the first author
 \centerline{Center for Nanoscience and Technology, Italian Institute of Technology, Polytechnic University of Milan, 20133 Milan, Italy}
} % Do not forget to end the {\footnotesize by the sign }

\medskip

\centerline{\scshape Ville Kolehmainen}
\medskip
{\footnotesize
% please put the address of the first author
 \centerline{Department of Applied Physics, University of Eastern Finland, P.O. Box 1627, 70211 Kuopio, Finland}
} % Do not forget to end the {\footnotesize by the sign }

\bigskip

% The name of the associate editor will be entered by an editorial staff
% "Communicated by the associate editor name" is not needed for special issue.
 \centerline{(Communicated by the associate editor name)}

%The abstract of your paper
\begin{abstract}
In fluorescence diffuse optical tomography (fDOT), the reconstruction of the fluorophore 
concentration inside the target body is usually carried out using a normalized Born approximation model where
the measured fluorescent emission data is scaled by measured excitation data. One of the benefits of the model is that it can tolerate inaccuracy in the absorption and scattering distributions that are used in the construction of the forward model to some extent. In this paper, we employ the recently proposed Bayesian approximation error approach to fDOT for compensating for the 
modeling errors caused by the inaccurately known optical properties of the target in combination with the normalized Born approximation model. The approach is evaluated using a simulated test case with 
different amount of error in the optical properties. The results show that the Bayesian approximation error approach improves 
the tolerance of fDOT imaging against modeling errors caused by inaccurately known absorption and scattering of the target. 
\end{abstract}

\section{Introduction}
\label{sect:intro}  % \label{} allows reference to this section

Fluorescence diffuse optical tomography (fDOT) is an emerging imaging technique aiming at recovering the distribution of fluorophore marker inside diffusive target medium from measurements of fluorescent emission at the surface of the body \cite{Yodh,Ntz05}. Typically, fluorescence agents bound to molecules or proteins are introduced to the bloodstream. The molecules then act as ligands as they attach themselves to the targeted receptor sites. The surface of the body is illuminated with light at the excitation wavelength of the fluorescence agent. The measurement system collects fluence from the body surface at the emission wavelength. The inverse problem associated with fDOT is the estimation of spatially distributed fluorophore marker concentration in the body.

In small animal studies \cite{Graves03,Ntz05,ntzi2005}, fDOT has been used for monitoring tumors in brain \cite{Ntz2002, Davis2010}, breast \cite{Patwardhan05, Shih2006} and lungs \cite{Josserand2008}. fDOT has also been used to monitor brain strokes \cite{Martin2008}, localizing lymph nodes that get affected at the onset of cancer \cite{Solomon2011} and studying the effects of drugs on tumor \cite{Ntz2004}. In humans, fDOT has been used in imaging breast cancer \cite{Yodh,Yodh1,Ven2010,Erickson2013}.

Computationally, the inverse problem in fDOT amounts to estimating the spatially distributed fluorophore concentration from the model
\be \label{cemmod1}
y = A(\mua,\mus) h + e,
\ee
where $h \in \R^n$ is the vector of unknowns, representing a pixel or voxel or other parametrization of the  
fluorophore concentration, $y \in \R^\nmeas$ is the measurement vector, 
$e \in \R^\nmeas$ models the measurement noise and $A(\mua,\mus)$ is matrix implementing the forward model (i.e., the light propagation model) corresponding to nominal absorption and scattering distributions $\mua$ and $\mus$. The estimation of $h$ from the model
(\ref{cemmod1}) is an ill-posed inverse problem, that is sensitive with respect to measurement and modeling errors.

The fDOT inverse problem is most often carried out using the so-called normalized Born approximation model
\cite{ntzi2001}, where the measurement vector is the measured 
fluorescent emission data vector $\yfo$ scaled by the measured excitation data $\yeo$ as
\be \label{nborn1}
y = \frac{\yfo}{\yeo},
\ee 
and forward matrix $A$ is 
\be \label{nborn2}
A(\mua,\mus) = {\rm diag}(\frac{1}{\yec}) \tilde A (\mua,\mus)
\ee
where $\tilde A$ is the forward matrix corresponding to the raw fluorescence measurement and 
$\yec$ is computed excitation data corresponding to the absorption and scattering distributions  $\mua$ and $\mus$ \cite{ntzi2001,rudge2010}. The convenience of the Born normalization comes from the fact that it does not require
a reference excitation measurement from homogeneous reference media. The normalization also effectively calibrates the
problem with respect to source strength and individual gains and coupling coefficients of individual source detector pairs \cite{ntzi2001,ntzi2005}. From the practical point of view, a further significant feature of the Born normalized model is
that it can tolerate inaccurately known target absorption and scattering distributions $(\mua,\mus)$ to some extent. 

The absorption and scattering distributions $(\mua,\mus)$ are often interpreted as known (nuisance) parameters in the fDOT problem. 
However, in practical experiments, the actual values of these parameters are usually not known accurately, and therefore
one is bound to use approximate measurement model 
\be \label{cemmod2}
y \approx A(\muabar,\musbar) h + e,
\ee
where $\muabar$ and $\musbar$ are our estimates for the absorption and scattering of the body.
 Obviously, the use of incorrect realizations $(\muabar,\musbar)$ in the measurement model 
induces unknown modeling error $$[A(\mua,\mus) - A(\muabar,\musbar)]h$$ in the model, 
and since the inverse problem is ill-posed, this error may cause large artifacts to the reconstructed 
fluorophore image.

In this paper, we propose the reduction of the reconstruction errors caused by inaccurately known absorption and scattering of the body by the Bayesian approximation error approach \cite{kaipio05,kaipio07}. We consider the approach using the Born normalized model (\ref{cemmod1}-\ref{nborn2}), so that the starting point would be the model that has the best available tolerance for the inaccurately known absorption and scattering. 
 The key idea in the Bayesian approximation error model is to represent not just the measurement 
error, but also computational model inaccuracy as a random variable. Hence, instead 
of the approximate measurement model (\ref{cemmod2}), we write an {\em accurate} measurement model,
\bea 
y &=& A(\muabar,\musbar) h + \underbrace{\left[ A(\mua,\mus)h - A(\muabar,\musbar)h \right]}_{\ape (\mua,\mus,h)} + e \\
&=&  A(\muabar,\musbar) h + \nu \label{aemmod1} 
\eea
where the term $\ape = [A(\mua,\mus) - A(\muabar,\musbar)]h$ represents
the approximation error and $\nu = \ape + e$ the total noise.
%Note that the model (\ref{aemmod1}) is exact, 
% see equation (\ref{cemmod1}). 
Obviously, the realization of $\ape$
is unknown since it depends on the unknown $h$ and the unknown 
nuisance parameters $(\mua,\mus)$. However, in the Bayesian inversion
paradigm, we can compute an estimate for the statistics of the approximation error $\ape$ 
over the prior probability distributions of the unknowns and the nuisance parameters
$\mua$ and $\mus$. 
The approximation error statistics are then used in the inverse problem
to compensate for the inaccurately known $\mua$ and $\mus$.

The Bayesian approximation error approach was originally 
applied for discretization error in several different applications in Ref.~\cite{kaipio05}.
For this reason, the term ``approximation error'' is commonly used 
also where ``modeling error'' might be a more appropriate term.
The approach was verified with real EIT data in Ref.~\cite{nissinen2008}, 
where the approach was employed for the compensation of discretization errors 
and the errors caused by inaccurately known height of the air-liquid surface in an industrial mixing tank.
The application of the Bayesian approximation error approach for 
the discretization errors and the truncation of
the computational domain was studied in Ref.~\cite{lehikoinen07}, and for the
linearization error in Ref.~\cite{pursiainen06}. In Ref.~\cite{nissinen2009} the approach was
evaluated for the compensation of errors caused by 
coarse discretization, domain truncation and 
unknown contact impedances with real EIT data. 
In addition to EIT, the Bayesian approximation error approach has also been applied 
to other inverse problems and other types of (modeling) errors:
Model reduction, domain truncation and unknown anisotropy structures in optical diffusion
tomography were treated in Refs.~\cite{arridge06}, \cite{Kolehmainen2009}, \cite{heino04} and \cite{Heino2005}.
Missing boundary data in the case of image processing 
was considered in Ref.~\cite{calvetti2005}.
In Ref.~\cite{Tarvainen2010}, again related to optical tomography,
an approximative physical model (diffusion model instead of the radiative transfer model) was used for the forward problem.
In Ref.~\cite{Kolehmainen2011}, an unknown nuisance
distributed parameter (scattering coefficient)
was treated with the Bayesian approximation error approach.
The compensation of errors caused by unknown optode coupling coefficients and locations was considered in Ref.~\cite{Mozumder2013}. The compensation of errors caused by unknown domain shape and discretization error was considered in Ref.~\cite{Mozumder2014}. The extension and application of the modeling error approach to time-dependent
inverse problems was considered in Refs.~\cite{jhuttunen2007a}, \cite{jhuttunen2007b} and \cite{voutilainen09}.

In this work, we evaluate the Bayesian approximation error approach with simulated two dimensional (2D) and three dimensional (3D) examples. In the 2D example, we use different levels of severity of error in the nominal absorption and scattering distributions and show that the approach improves the tolerance of the fDOT problem against inaccurately known absorption and scattering of the body over the conventional reconstruction approach. In the 3D example, we test the approach using simulated data using the Digamous atlas geometry \cite{Dogdas2007,DigimouseWeb}.

The remainder of the paper is organized as follows. In Section \ref{sect:AEmodel1}, the Bayesian approximation error model for the compensation of modeling errors due to unknown absorption and scattering in fDOT is presented. In Section \ref{sect:fwdmod}, we review the light transport model which we use for the forward model. The details of data simulation, meshing, constructing the prior models and estimation of the approximation error statistics are described in Section \ref{sect:Comp}. The simulation results are presented in Section \ref{sect:Results}. Finally, the conclusions are given in Section \ref{sect:Discussion}.

% and elements of $\tilde A$ are of the form
% \[
%\tilde A_{(\ell-1)\ndet + j ,k} = \int_\domain \Psi_j^\ast (\posvec) \chi_k (\posvec) \Phi_\ell (\posvec) \rmd \posvec , \quad \ell = 1 \ldots \nsource, \quad j = 1,\ldots,\ndet., 
% \]
% with $\chi_k$ denoting the characteristic function of image pixel (or voxel) $k$, see \cite{rudge2010}. 

\section{Bayesian approximation error approach}
\label{sect:AEmodel1}

\subsection{Statistical inversion in general}
\label{sect:StatInv}

In the Bayesian approach to inverse problems, the principle is that all unknowns and measured quantities are considered as random variables and the uncertainty of their values are encoded into their probability distribution models \cite{kaipio05,calvetti2007}. The complete model of the inverse problem is the posterior density
model, that is, the information and uncertainities in the unknown or interesting variables given the measurements, given by the Bayes' theorem
\begin{equation} \label{post1}
\pi(h,\mua,\mus,e|y)=\frac{\pi(y|h,\mua,\mus,e)\pi(h,\mua,\mus,e)}{\pi(y)},
\end{equation}
where $\pi(y|h,\mua,\mus,e)$ is the likelihood density modeling the probability of different measurement realizations when the realizations of $h$, $\mua$, $\mus$ and $e$ are given. The density $\pi(h,\mua,\mus,e)$ is the prior model and it models our information on the unknown parameters before the actual measurements. The posterior (\ref{post1}) is practically always marginalized with respect to the unknown but uninteresting measurement related errors $e$ as
\begin{equation} \label{post2}
\pi(h,\mua,\mus|y) = \int \pi(h,\mua,\mus, e \vert y) \rmd e,
\end{equation}
for details in the case of the additive error model $y = A(h) + e$, see Refs.~\cite{Kolehmainen2011} and \cite{Bay13}, for other types of errors see Ref.~\cite{kaipio05}. The posterior density $\pi(h,\mua,\mus|y)$ is a probability density in a very high-dimensional space. Thus, in order to get practical estimates for the unknowns and visualize the solution, one needs to compute point estimate(s) from the posterior density, the most common choice in high dimensional cases being the {\em maximum a posteriori} (MAP) estimate. In principle, one
could attempt to compute the MAP estimate for all the unknown model parameters
\begin{equation} \label{mapall}
(h,\mua,\mus)_{\rm MAP} = \arg \max_{h,\mua,\mus} \pi (h,\mua,\mus \vert y).
\end{equation} 
See Refs.~\cite{Tan2008}, \cite{Soloviev2009} and \cite{Lin2009} for simultaneous reconstruction of fluorescent and optical parameters. Also, in Ref.~\cite{Correia2013} a method to determine a modified optical parameter (related to $\mua$, $\mus$) from measured excitation data and include it as apriori anatomical information into fDOT imaging is presented. Alternatively, one could treat the uncertainty in the values of nuisance parameters $(\mua,\mus)$ by marginalizing the posterior density  as
\begin{equation} \label{marg1}
\pi (h \vert y) = \int \int \pi (h,\mua,\mus \vert y) \rmd \mua \rmd \mus
\end{equation} 
and then compute estimate for the primary unknowns from the posterior $\pi (h \vert y)$.
However, the solution of $(\ref{marg1})$ would require Markov chain Monte Carlo integrations that would be computationally infeasible for practical purposes. 

The key idea in the Bayesian approximation error approach is to find an approximation $\tilde \pi (h \vert y)$ for the posterior (\ref{marg1}) such that the marginalization over the uncertainty in the values of $(\mua, \mus)$ is carried out {\em approximately} 
and in a computationally feasible way.  

Before presenting the Bayesian approximation error approach for treating the uncertainty in the optical parameters $(\mua, \mus)$, we
first review the standard fDOT reconstruction approach where $(\mua, \mus) = (\muabar,\musbar) $ are treated as known and fixed variables.

\subsection{Conventional error model}
\label{sect:CEmodel}

In most of fDOT reconstruction schemes, the optical parameters $(\mua, \mus)$  are treated as known (fixed) nuisance parameters with values $\mua = \muabar$ and $\mus = \musbar$. Within the Bayesian setup, this is equivalent to considering
$(\mua, \mus)$ as fixed conditioning parameters, leading to
posterior density model
\be\label{mm}
\pi(h,e|y,\mua=\muabar,\mus=\musbar)=\frac{\pi(y|h,\mua=\muabar,\mus=\musbar,e)\pi(h)\pi(e)}{\pi(y)}.
\ee
However, in cases that the values $(\muabar, \musbar)$ are incorrect, the model (\ref{mm}) can be grossly misleading.

Given the observation model (\ref{cemmod2}) with fixed realizations $(\mua, \mus) = (\muabar,\musbar) $, and modeling the measurement noise as normal $ e \sim \mathcal{N}(e_*,\Gamma_e)$, where $e_* \in \R^\nmeas$ is the measurement noise mean and $\Gamma_e \in \R^{\nmeas \times \nmeas}$ is the measurement noise covariance and marginalizing (\ref{mm}) over the unknown measurement errors $e$ 
as $$ \pi(h \vert y,\muabar,\musbar) = \int \pi(h,e|y,\muabar,\musbar) \rmd e$$
the posterior density becomes \cite{Kolehmainen2011,Bay13}
\be\label{postcem}
\pi(h \vert y,\muabar,\musbar) \propto  \exp\left\{-\frac{1}{2}
\| y - A(\muabar,\musbar) h - e_* \|^2_{\Gamma^{-1}_{e}} \right\} \pi(h)
\ee  
In addition, if the random measurement noise has zero mean ($e_*=0$), the MAP estimate corresponding to the posterior (\ref{postcem}) is obtained as  
\begin{eqnarray}\label{mapcem0}
 \hcem & =& \arg\hspace{1mm}\max_h\hspace{1mm}\pi (h|y,\mua=\muabar,\mus=\musbar)    \nonumber \\
   & = & \arg\hspace{1mm}\min_{h}\hspace{1mm}\{ \| y-A(\muabar,\musbar) h \|^2_{\Gamma^{-1}_{e}}  - 2 \log\pi(h)\}, \label{mapcem}
\end{eqnarray}
where the Cholesky factor $L_e^{\rm T}L_e=\Gamma_e^{-1}$. 
We refer to the solution of (\ref{mapcem0}) as the MAP estimate with the conventional error model (CEM) approach.

\subsection{Bayesian approximation error model}
\label{sect:AEmodel2}

In the Bayesian approximation error approach, we write the measurement model as
\bea 
y &=& A(\muabar,\musbar) h + \underbrace{\left[ A(\mua,\mus)h - A(\muabar,\musbar)h \right]}_{\ape (\mua,\mus,h)} + e \\
&=&  A(\muabar,\musbar) h + \nu \label{aemmod2} 
\eea

Note that the model (\ref{aemmod2}) is exact, 
see equation (\ref{cemmod1}). Here the term $\ape = [A(\mua,\mus) - A(\muabar,\musbar)]h$ represents
the approximation error and $\nu = \ape + e$ the total error. Obviously, the realization of $\ape$
is unknown since it depends on the unknowns $h$ and the unknown 
nuisance parameters $(\mua,\mus)$. However, in the Bayesian inversion
paradigm we can compute an estimate of the statistics of $\ape$ using the
prior probability distributions of the unknowns and the nuisance parameters and approximately marginalize over $\ape$.

To include the uncertainty in the noise process $\ape$ into our computational posterior model, the core step in the Bayesian approximation error approach is approximate pre-marginalization of the joint distribution of the parameters $(y,h,e,\ape,\mua,\mus)$ over the nuisance parameters $(e,\ape,\mua,\mus)$. Following the approach in Refs.~\cite{Kolehmainen2011} and \cite{Bay13} and making a Gaussian approximation for the joint density $\pi (h,\ape)$, we obtain the approximate 
likelihood model
\be
\tilde \pi(y \vert h) \propto \exp\left\{-\frac{1}{2}
\| y - A(\muabar,\musbar) h - \nu_{\ast \vert h} \|^2_{\Gamma^{-1}_{\nu \vert h}} \right\}
\ee
where
\bea
\nu_{\ast \vert h} &=& e_\ast + \ape_\ast + \Gamma_{\ape,h} \Gamma_h^{-1} (h- h_\ast) \label{numean} \\
\Gamma_{\nu \vert h} &=& \Gamma_e + \Gamma_\ape - \Gamma_{\ape,h} \Gamma^{-1}_h \Gamma_{h,\ape} \label{nucov} 
\eea

For the posterior model, we write an approximation 
$\tilde \pi (h \vert y) \propto \tilde \pi (y \vert h) \pi(h)$, leading
to MAP estimation problem
\be \label{mapaem}
\haem = \arg \min_h \{ \| y - A(\muabar,\musbar) h - \nu_{\ast \vert h} \|^2_{\Gamma^{-1}_{\nu \vert h}} - 2 \log \pi (h) \}. 
\ee
We refer to the estimate (\ref{mapaem}) as the MAP with the Bayesian approximation error model (MAP-AEM).

In the following, we employ the {\em enhanced error model} \cite{kaipio05,kaipio07}, where we approximate the error $\ape$ and the unknown $h$ as mutually uncorrelated, implying that the mean and covariance in equations (\ref{numean})-(\ref{nucov}) become
\[
\nu_{\ast \vert h} \approx e_\ast + \ape_\ast, \quad \Gamma_{\nu \vert h} 
\approx \Gamma_e + \Gamma_\ape.
\]
In this work, the mean $\ape_\ast$ and covariance $\Gamma_\ape$ were computed by sampling based Monte Carlo integration (see Section \ref{sect:SamplingErrors}).

\subsection{Estimates to be computed}
\label{sect:Est}

To evaluate the Bayesian approximation error approach, we compute the following estimates
\begin{description}
\item[(MAP-REF)] The unknown $\hcem$ using correct fixed optical coefficients $(\mua,\mus)$:
\be \label{mapref}
\hreff = \arg \min_h \{ \| y - A(\mua,\mus) h \|^2_{\Gamma^{-1}_{e}} - 2 \log \pi (h) \}.
\ee
In other words, in MAP-REF $(\mua,\mus)$ are known exactly. Therefore, this estimate serves as reference of conventional reconstruction when 
$(\mua,\mus)$ are known.

\item[(MAP-CEM)] The unknown $\hcem$ using incorrect fixed optical coefficients $(\muabar,\musbar)$ (Eq. \ref{mapcem0}). This estimate serves as a conventional reconstruction when nominal $(\mua,\mus)$ are incorrect. 

\item[(MAP-AEM)] The unknown $\haem$ using the same fixed incorrect coefficients $(\muabar,\musbar)$ (Eq. \ref{mapaem}). This estimate serves as a Bayesian approximation error model reconstruction when nominal $(\mua,\mus)$ are incorrect.

\end{description}

The values $(\muabar,\musbar)$ in the computations correspond to the expectations of the (proper Gaussian smoothness) prior models $\pi (\mua)$ and $\pi(\mus)$, respectively. The prior means were homogeneous distributions with $\muabar(r) \equiv 0.01 \coefunit$ and $\musbar(r) \equiv 1 \coefunit$.

%In the following, we compute the MAP estimates corresponding to proper Gaussian smoothness prior model $\pig$. 

\section{Forward model}
\label{sect:fwdmod}

Let $\Omega \subset \mathbb{R}^{n}$, $n$ = 2,3, denote the object domain. In a diffusive medium like soft tissue, the commonly used light transport model for excitation and fluorescence light is the diffusion approximation (DA) to the radiative transport equation (RTE) \cite{Ishimaru}. In this paper the DC (zero-frequency) domain version of the diffusion approximation is used 
\be\label{deeqn1}
 \left(-\nabla \cdot \kappa(r) \nabla  + \mua(r) \right) \Phi^{\rm e}(r) = 0, \hspace{0.5cm} r \in \Omega, 
\ee
\be\label{deeqn2}
\Phi^{\rm e}(r)+\frac{1}{2\zeta}\kappa(r) \alpha \frac{\partial \Phi^{\rm e}(r)}{\partial \vartheta} = \left\{ \begin{array}{ll}
         \frac{q(r)}{\zeta} & r \in r_s\\
         0 & r \in \partial \Omega \setminus r_s \end{array} \right. , 
\ee
\be\label{deeqn3}
 \left(-\nabla \cdot \kappa(r) \nabla  + \mua(r) \right) \Phi^{\rm f}(r) = h(r)\Phi^{\rm e}(r), \hspace{0.5cm} r \in \Omega, 
\ee
\be\label{deeqn4}
\Phi^{\rm f}(r)+\frac{1}{2\zeta}\kappa(r) \alpha \frac{\partial \Phi^{\rm f}(r)}{\partial \vartheta} = 0,\hspace{0.5cm} r \in \partial \Omega,
\ee
where $\Phi^{\rm e}(r) := \Phi^{\rm e}$ is the excitation photon density, $\Phi^{\rm f}(r) := \Phi^{\rm f}$ is the fluorescence emission photon density, $\mua(r) := \mua$ is the absorption coefficient, $\kappa(r) := \kappa$  is the diffusion coefficient. The diffusion coefficient $\kappa$ is given by $\kappa(r) = 1/(n(\mua(r)+\mus(r)))$, where $\mus(r) := \mus$ is the reduced scattering coefficient. For simplicity the spectral dependency of the optical properties $(\mua, \mus)$ is omitted and they are modelled the same at the excitation and emission wavelengths. The parameter $h(r) := h$ is the fluorophore concentration. The parameter $q(r)$ is the strength of the light source at location $r_s \subset \partial\Omega$. The parameter $\zeta$ is a dimension dependent constant ($\zeta$ = $1/\pi$ when $\Omega \subset \mathbb{R}^{2}$, $\zeta$ = $1/2$ when $\Omega \subset \mathbb{R}^{3}$), $\alpha$ is a parameter governing the internal reflection at the boundary $\partial \Omega$ and $\vartheta$ is the outward normal to the boundary at point $r$. The measurable excitation data and fluorescence emission data (Eq. \ref{nborn1}) are given by
\bea
y^{\rm e}(r) & = &  -\kappa \frac{\partial \Phi^{\rm e}(r)}{\partial \vartheta} = \frac{2\gamma}{\alpha}\Phi^{\rm e}(r), \hspace{6mm} r \in r_d, \label{bdd1} \\
y^{\rm f}(r) & = & -\kappa \frac{\partial \Phi^{\rm f}(r)}{\partial \vartheta} = \frac{2\gamma}{\alpha}\Phi^{\rm f}(r), \hspace{6mm} r \in r_d, \label{bdd2}
\eea
where $r_d  \subset \partial\Omega$ are the detector locations. 

For the inverse problem, the mapping $A(\mua, \mus)h$ (Eq. (\ref{cemmod1})) is given by \cite{ntzi2001,ntzi2005,rudge2010}
\be\label{jacobian}
A(\mua, \mus)h = \frac{\int_\Omega \Phi^{\rm e}(r_s,r) \Psi^{\rm e}(r_d,r) h(r) \rmd r}{\int_\Omega \Phi^{\rm e}(r_s,r) \rmd r},   
\ee
where $\Phi^{\rm e}(r_s,r)$ is the computed excitation photon density due to source $q(r)$. $\Psi^{\rm e}(r_d,r)$ is the computed adjoint solution (photon density due to sources placed at detector locations $r_d$).

The numerical approximation of the forward model used here is based on a finite element method (FEM) solution of Eq. (\ref{deeqn1}-\ref{bdd2}).  

\section{Computation details}
\label{sect:Comp}

\subsection{Simulation of the measurement data}

\subsubsection{2D simulations}

In the 2D numerical studies, the domain $\Omega \subset \mathbb{R}^2$ was a disk with radius $r = 25 \mm$. The measurement setup consisted of 16 sources and 16 detectors. The source and detector optodes were modeled as $1 \mm$ wide surface patches located at equi-spaced angular intervals on the boundary $\partial\Omega$. With this setup, the vector of fDOT measurements (\ref{nborn1}) was $y \in \mathbb{R}^{256}$. Five targets with different optical properties of absorption and scattering were used in the simulations (see Fig. \ref{case1opt}, column 1, 2). The fluorophore concentrations of these targets is shown in  column 1, Fig. \ref{case1}. For the simulation of the measurement data, a mesh with 33806 nodes and 67098 triangular elements was used. In the inverse problem, a FEM mesh with 26075 nodes and 51636 elements was used.

\subsubsection{3D simulations}

In the 3D simulations, the domain $\Omega \subset \mathbb{R}^3$ was a mouse atlas ``Digimouse'' \cite{Dogdas2007,DigimouseWeb}. The measurement setup consisted of 32 sources and 32 detectors (see Fig. \ref{case2}, bottom right). The source and detector optodes were modeled as $1 \mm$ wide surface patches placed on the top-surface of the boundary $\partial\Omega$. With this setup, the vector of fDOT measurements (\ref{nborn1}) was $y \in \mathbb{R}^{512}$. The optical properties of the target are shown in column 1, Fig. \ref{case2}. The fluorophore concentrations of the target is shown in column 2, Fig. \ref{case2}. For generating the measurement data and for the inverse problem we used the same mesh obtained from the Digimouse website \cite{DigimouseWeb} which had 58244 nodes and 306773 elements.\\\

The simulated measurement data was generated using the FEM approximation of Eq (\ref{deeqn1})-(\ref{bdd2}). Random measurement noise, drawn from a zero-mean Gaussian distribution was added separately to the simulated measurement excitation and fluorescence emission data as,
\bea 
\frac{y^{\rm f}_{\rm obs}}{y^{\rm e}_{\rm obs}} &=& \frac{y_{\rm calc}^{\rm f} + e^{\rm f}}{y_{\rm calc}^{\rm e} + e^{\rm e}} \label{noisyborn} \\
&=& y_{\rm calc} + e.
\eea
Here $e^{\rm f} \sim \mathcal{N} (0, {\rm diag}(\sigma_{e^{\rm f},1},..,\sigma_{e^{\rm f},m})) \in \mathbb{R}^m $ is the noise added to the fluorescence emission data, $e^{\rm e}\sim \mathcal{N} (0, {\rm diag}(\sigma_{e^{\rm e},1},..,\sigma_{e^{\rm e},m})) \in \mathbb{R}^m$ is the noise added to the excitation data. The standard deviations $\{\sigma_{e^{\rm e},1},..,\sigma_{e^{\rm e},m}\}$ and $\{\sigma_{e^{\rm f},1},..,\sigma_{e^{\rm f},m}\}$ were specified as $1\%$ of the simulated noise free measurement excitation and fluorescence emission data, $y_{\rm calc}^{\rm e}$ and $y_{\rm calc}^{\rm f}$ respectively. 

In order to estimate the noise statistics for the inverse problem, a Gaussian approximation for the measurement noise $e$, $\pi (e) = \mathcal{N}(0, \Gamma_e)$ was constructed. The covariance $\Gamma_e$ was approximated by a diagonal 
where the diagonal elements (i.e., standard deviations of the measurements) were estimated as sample averages from 
%the noise were obtained by creating 
100 noisy realizations of the Born ratio (Eq. (\ref{noisyborn})). In a practical setup, similar estimation of $\Gamma_e$ can
be carried out by repeated measurements from a phantom target.

\subsection{Prior models}
\label{sect:Prior}

For drawing the samples of optical coefficients $x$, where $$x=(\mua,\mus, h)^{\rm T},$$ for the estimation of 
the approximation error statistics, we used a Gaussian (Markov random field) \cite{Havard2005} prior model along with non-negativity constraint \cite{Kolehmainen2006},
\be
\pi(x) = \pi_{\rm G}(x)\pi_+(x).
\ee
Here $\pi_{\rm G}(x)$ is a Gaussian smoothness prior model and $\pi_+(x)$ is the non-negativity constraint. Section \ref{sect:GPrior} describes the implementation of the Gaussian smoothness prior model, Section \ref{sect:PosPrior} describes the implementation of non-negativity constraint.

\subsubsection{Proper Gaussian smoothness prior}
\label{sect:GPrior}

As we need to draw samples of the unknown $x$ for the estimation of approximation errors, we need a proper (integrable) prior distribution for the unknowns. In this study we used a proper Gaussian smoothness prior $\pi_{\rm G}(x)$ for the unknowns. In this model, the absorption, scattering and fluorophore concentration images $\mua$, $\mus$ and $h$ were modeled 
as mutually independent Gaussian random fields with a 
joint prior model 
\begin{equation} \label{smoothnessprior}  
\pi_{\rm G} (x)  \propto \exp\{ -\frac{1}{2} \| L_{x} (x - x_* )\|^2 \}, \quad  L_{x} \transp L_{x} = \Gamma_{x}^{-1}
\end{equation}
where
\[
x_* = \mtrx{c}{\muabar \\ \musbar \\ h_*}, \quad \Gamma_{x} = \mtrx{ccc}{\Gamma_{\mua}  & 0 & 0 \\ 0 & \Gamma_{\mus} & 0 \\ 0 & 0 & \Gamma_h}. 
\]
In the construction of the mean vectors $\muabar,\musbar,h_*$ and covariances $\Gamma_{\mua}$, $\Gamma_{\mus}$ and $\Gamma_h$,
the random field, say $f$ (i.e., either $\mua$, $\mus$ or $h$), is 
considered in the form
\[
f = f_{\rm in} + f_{\rm bg}
\]
where $f_{\rm in}$ is a spatially inhomogeneous  
parameter with zero mean, 
$$
f_{\rm in} \sim \mathcal{N}(0,\Gamma_{{\rm in},f})
$$
and $f_{\rm bg}$ is a spatially constant (background)
parameter with non-zero mean.
For the latter, we can write $f_{\rm bg}  = q\mathbb{I}$, where $\mathbb{I} \in \mathbb{R}^n$ is a 
vector of ones and $q$ is a scalar random variable with distribution
$q\sim\mathcal{N}(c,\sigma_{{\rm bg},f}^2)$.
In the construction of $\Gamma_{{\rm in},f}$, the approximate correlation length
can be adjusted to match the size of the expected inhomogeneities and the marginal variances
of $f_k$:s are tuned based on the expected contrast of the inclusions.
We model the distributions $f_{\rm in}$ and $ f_{\rm bg}$ 
as mutually independent,
that is, the background is mutually independent with the inhomogeneities.
Thus, we have  
$$f_* = c \mathbb{I},\quad  \Gamma_f = \Gamma_{{\rm in},f} + \sigma_{{\rm bg},f}^2 \mathbb{I}\mathbb{I}\transp$$ 
See \cite{kaipio05,arridge06,Kolehmainen2011} for further details, and see
\cite{Lieberman2010} for an alternative construction of a proper smoothness prior.

The parameters in the prior model $\pi (x)$ were selected as follows.
The mean for background absorption, scattering and fluorophore concentration were set as $\submua{\ast} = 0.01 \coefunit$, $\submus{\ast} = 1 \coefunit$ and $h_* = 0 \coefunit$.
The standard deviations $\sigma_{{\rm bg},\mua}$ and $\sigma_{{\rm bg},\mus}$ of the background values were chosen such that 2 s.t.d. limits equaled $25 \%$ of the mean values $\submua{\ast}$ and $\submus{\ast}$ and the standard deviation $\sigma_{{\rm bg},h}$ of the background value for $h$ was chosen as 0.25 $\coefunit$. In the construction of $\Gamma_{{\rm in },f}$, 
the correlation length for $\mua$, $\mus$ and $h$ in the prior was set as 16mm. The marginal standard deviations were set to equal values in each pixel and $\sigma_{{\rm in},\mua}$ and $\sigma_{{\rm in},\mus}$ were chosen such that 2 s.t.d. limits equaled $50 \%$ of the mean values $\submua{\ast}$ and $\submus{\ast}$. $\sigma_{{\rm in},h}$ was chosen as 1$\coefunit$. Thus, the overall marginal variances (i.e., diagonal elements of $\Gamma_{\mua}$, $\Gamma_{\mus}$ and $\Gamma_h$) were $2\sigma_{\mua} = 0.0056 \coefunit$, $2\sigma_{\mus} = 0.56 \coefunit$ and $2\sigma_h = 1 \coefunit$. This gives overall 2 s.t.d. intervals $\mua \in [0.0044, 0.0156] \coefunit$, $\mus \in [0.44, 1.56] \coefunit$ and $h \in [-1, 1]$ i.e., the values of absorption, scattering and fluorophore concentration are expected  to lie within theses intervals with {\em prior} probability of $95\%$. The same prior parameters (correlation length, means and standard deviations) are used in 2D and 3D priors.

\subsubsection{Non-negativity constraint}
\label{sect:PosPrior}

In addition to the Gaussian smoothness prior, a non-negativity constraint was applied while drawing samples for the computation of approximation error statistics. A tolerance value for the values of $x$ was chosen and the values of $x$ drawn from prior $\pi_{\rm G}(x)$ that were less than the tolerance value, were set equal to the tolerance value as, 
\be
x = {\rm max}(x,{\rm tol}).
\ee
Here ``tol'' is the tolerance value, tol = 10e$^{-6}$ was used in this study.

\begin{figure}[ht]
\centerline{\includegraphics[width=100mm]{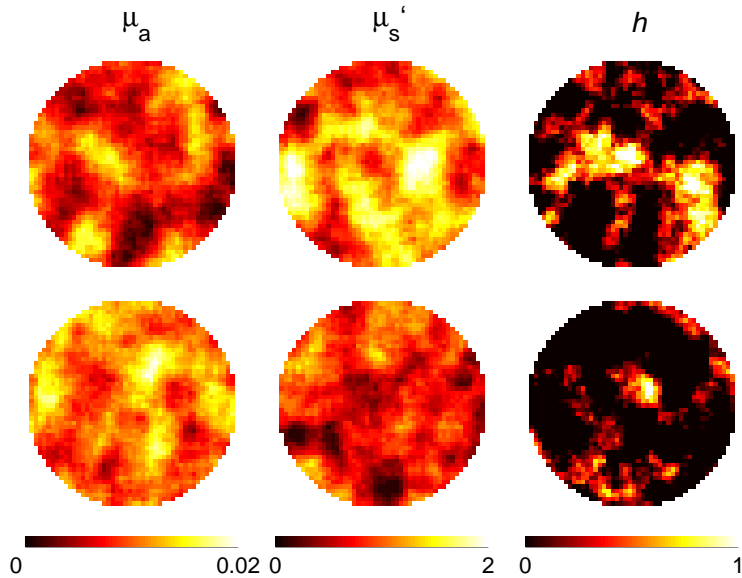}}
\caption{Two draws (top and bottom row) of $\mua$, $\mus$ and $h$ from $\pi(x)$ in the 2D simulation domain. From left: first column is $\mua$, second column is $\mus$ and third column is $h$.}
\label{case0}
\end{figure}

Two random draws of $ x =(\mua,\mus,h)^{\rm T}$ from $\pi(x)$ in the 2D simulation domain are shown in Figure \ref{case0}.

The non-negativity prior during the computation of the MAP estimates (\ref{mapref}) , (\ref{mapcem}), (\ref{mapaem}) is taken into account by applying an exterior point search method \cite{Kolehmainen2006}. In this method, the non-negativity problem is approximated by a sequence of unconstrained problems
\bea\label{mapconst}
h_{\rm MAP}& = & \arg \min_h \{ - \log(\pi(h|y)) - \log(\pi(h)) \}, \\ 
& = & \arg \min_h \{ - 2 \log(\pi(h|y)) + \|L_h(h-h_*)\|^2+E^{j}(h) \} 
\eea
where $\pi(h|y)$ is the likelihood and $\pi(h)$ is the prior probability distribution. $E^{j}(h)$ is a penalty functional that is used to penalize the negative components of the solution $h$, the super index $j$ is used to denote the $j^{\rm th}$ problem in the sequence. The mean $h_* \in \mathbb{R}^n$ and Cholesky factor $L_h^TL_h = \Gamma_h^{-1} \in \mathbb{R}^{n \times n}$ is obtained from the Gaussian smoothness prior model $\pi_{\rm G}(x)$. In this study we employ a functional of the form 
\be
E^{j}(h) = \sum_{k = 1}^{\rm n}\gamma^j\phi(h_k)
\ee
where
\be
\phi(h_k) = \left\{ \begin{array}{ll}
         (h_k)^2, & h_k<0\\
         0, & {\rm otherwise,} \end{array} \right.
\ee
and ${\gamma^j,j = 1,2,..,{\rm M}}$ is a sequence of increasing positive penalty parameters. The exterior point methods guarantee the non-negativity of the solution only in the asymptotic limit $j\to\infty$. In this paper, we used $\{ \gamma^1 = 1,\gamma^2 = 10,\gamma^3 = 100 \}$ as the sequence of exterior point parameters. The incorporation of the non-negativity constraint to the MAP estimates (\ref{mapref}) , (\ref{mapcem}), (\ref{mapaem}), leads to non-linear minimization problems, which were solved by a Gauss-Newton algorithm with an explicit line search algorithm \cite{Schweiger2005}.

\subsection{Estimation of approximation error statistics}
\label{sect:SamplingErrors}

For the computation of the error statistics, first the samples of absorption, scattering and fluorophore concentrations
\begin{equation}
S = \{x^{(\ell)},\;l = 1,..,{\rm N_s}\}
\end{equation}
were drawn. Two of the samples $x^{(\ell)}$ are shown in Figure \ref{case0}.
The samples were then used for the computation of the accurate forward solutions $A(\mua^{(\ell)}, \mus^{(\ell)} )h^{(\ell)} $ and approximate forward solutions $A(\muabar, \musbar)h^{(\ell)} $, and the samples of the approximation error $$\varepsilon^{(\ell)} = [A(\mua^{(\ell)} ,\mus^{(\ell)} ) - A(\muabar,\musbar)]h^{(\ell)}, $$ were computed. Then the mean $\varepsilon_*$ and covariance $\Gamma_\varepsilon$ were estimated using the samples $\{ \varepsilon^{(\ell)} \} $ as
\begin{equation}\label{aemean}
{\varepsilon}_* = \frac{1}{{\rm N}_s}\sum_{\ell =1}^{{\rm N}_s} \varepsilon^{(\ell)} 
\end{equation}
\begin{equation}\label{aecov}
\Gamma_{\varepsilon} = \frac{1}{{{\rm N}_s}-1} \sum_{\ell =1}^{{\rm N}_s} (\varepsilon^{(\ell)}-\varepsilon_*)(\varepsilon^{(\ell)}-\varepsilon_*)^{\rm T}.
\end{equation} 

\section{Results}
\label{sect:Results}

\subsection{2D simulations}

\begin{figure}[ht]
\centerline{\includegraphics[width=120mm]{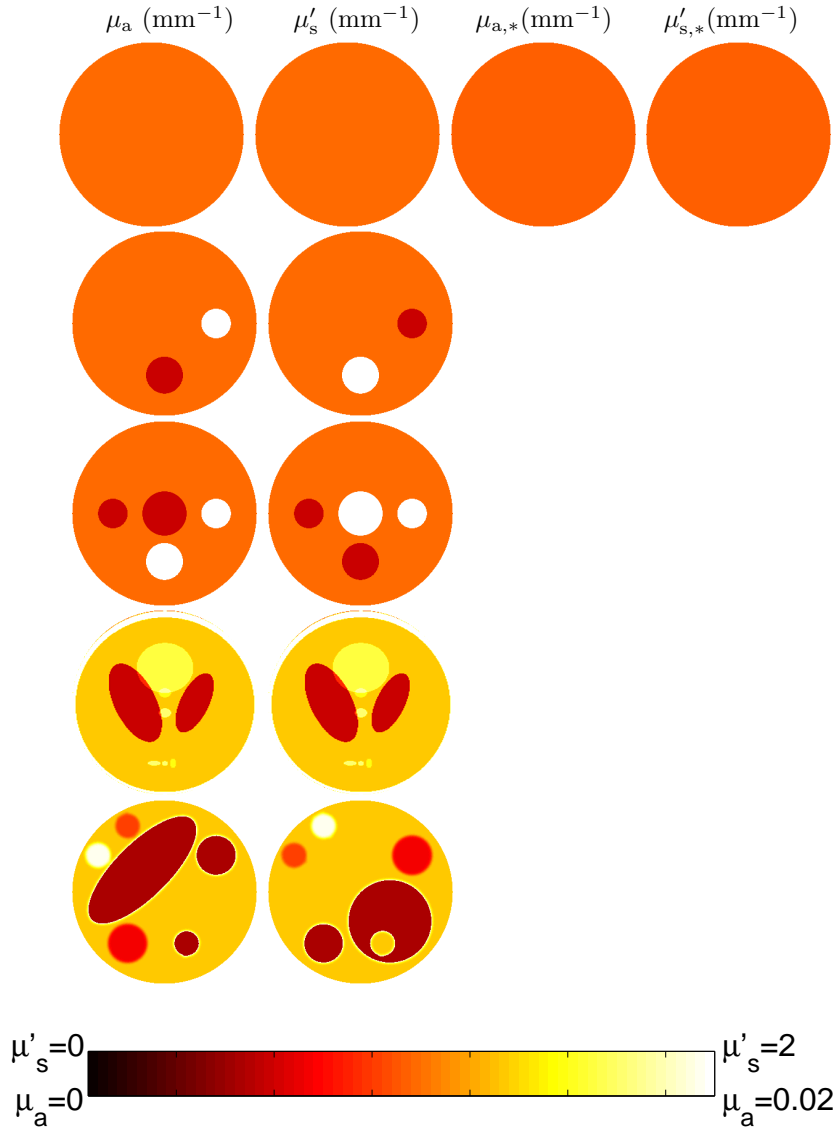}}
\caption{First and second columns show $\mua$ (left) and $\mus$ (right) of the body $\domain$ in test cases 1-5 (top to bottom). Third and fourth columns show the (incorrect) absorption and scattering $\muabar$ and $\musbar$ that are used the in the computation of the estimates MAP-CEM and MAP-AEM.}
\label{case1opt}
\end{figure}

\begin{figure}[ht]
\centerline{\includegraphics[width=100mm]{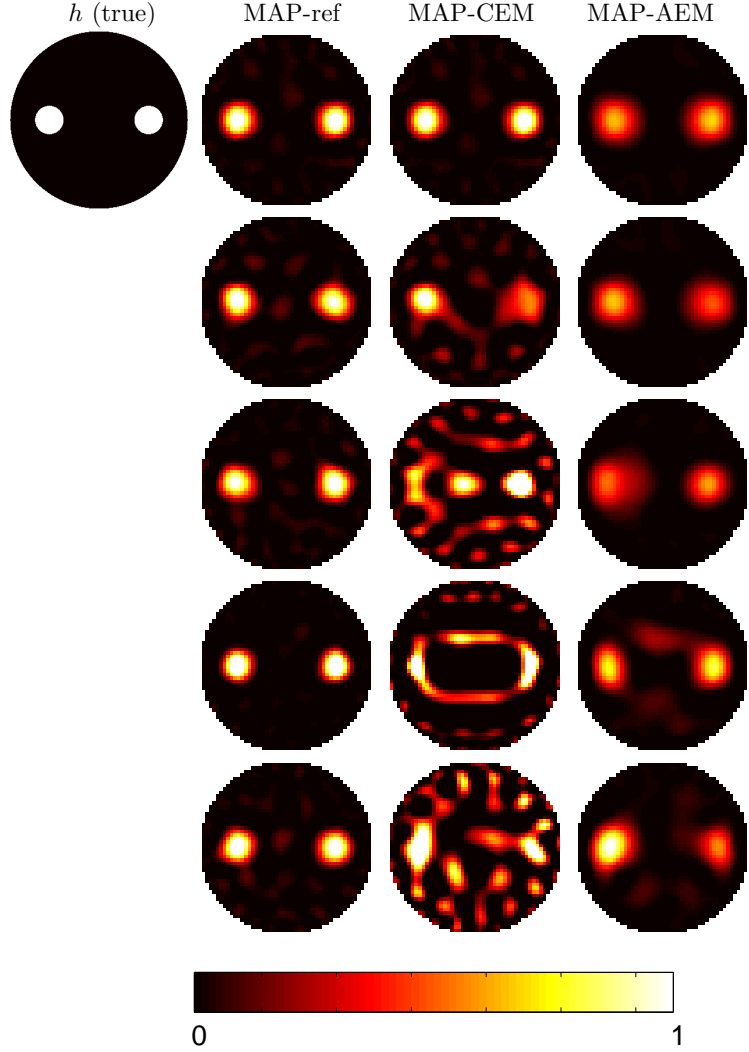}}
\caption{First column: fluorophore distribution $h$ of the body $\domain$ in test cases 1-5 (top to bottom). The second column shows the MAP-ref estimate using the correct absorption and scattering (forward matrix $A(\mua,\mus)$). The absorption and scattering images $\mua$ and $\mus$ are shown in columns 1 and 2 in Figure \ref{case1opt} (rows in respective order). Third column shows the the MAP-CEM estimate using the incorrect absorption and scattering values (forward matrix $A(\muabar,\musbar)$). The  $\muabar$ and $\musbar$ are shown in third and fourth column in Figure \ref{case1opt}. The fourth column shows the MAP-AEM estimates using the same incorrect forward matrix $A(\muabar,\musbar)$.}
\label{case1}
\end{figure}

The true absorption and scattering parameters $(\mua,\mus)$ used in the simulations
are shown in Figure \ref{case1opt}, column 1 (absorption) and column 2 (scattering). The nominal values $(\muabar, \musbar)$ that are used in the estimates $h_{\rm cem}$ and $h_{\rm aem}$  are shown in Figure \ref{case1opt}, column 3 (absorption), column 4 (scattering).  
%These values are equivalent to the background values of true $(\mua,\mus)$. 
The true target fluorophore distribution $h$ is shown in first column in Figure \ref{case1}. The MAP estimates with the measurement data from the target domains   
%($\mua$, $\mus$) corresponding to each row 
in columns 1 and 2 in Figure \ref{case1opt}
are shown in Figure \ref{case1}, column 2-4, in matching order of rows. The estimates are:

\begin{description}
\item[(MAP-REF)] The MAP-ref estimates using correct fixed $(\mua, \mus)$
\be\label{mapref1}
h_{\rm ref} = \arg \min_h \{ \| y - A(\mua,\mus) h \|^2_{\Gamma^{-1}_{e}} + \|L_h(h-h_*)\|^2+E^{j}(h) \}, 
\ee
are shown in column 2, Figure \ref{case1}. 
%The fixed values of the optical parameters 
%$(\mua, \mus)$ used in these estimates are the true target values of $(\mua, \mus)$. 
This corresponds to the reference estimate of conventional reconstruction when $(\mua,\mus)$ are known exactly.

\item[(MAP-CEM)] The MAP-CEM estimates with fixed optical coefficients $(\muabar, \musbar)$
\be\label{mapcem1}
h_{\rm cem} = \arg \min_h \{ \| y - A(\muabar,\musbar) h \|^2_{\Gamma^{-1}_{e}} + \|L_h(h-h_*)\|^2+E^{j}(h) \}, 
\ee
are shown in column 3, Figure \ref{case1}. This corresponds to estimate with conventional reconstruction when the nominal values of $(\mua,\mus)$ are incorrect. 

\item[(MAP-AEM)] The MAP-AEM estimates with the same fixed optical coefficients $(\muabar, \musbar)$
\be\label{mapaem1}
h_{\rm aem} = \arg \min_h \{ \| y - A(\muabar,\musbar) h - \nu_{\ast \vert h} \|^2_{\Gamma^{-1}_{\nu \vert h}} + \|L_h(h-h_*)\|^2+E^{j}(h) \}, 
\ee
are shown in column 4, Figure \ref{case1}. This corresponds to estimate with Bayesian approximation error 
model when the nominal values of $(\mua,\mus)$ are incorrect.

\end{description}

The relative error in the MAP estimates (\ref{mapref1}), (\ref{mapcem1}), (\ref{mapaem1}),
\be
{\rm Error} = \frac{\|h - h_{\rm true}\|^2}{\|h_{\rm true}\|^2} \times 100\%,
\ee
where $h$ is the estimated fluorophore distribution and $h_{\rm true}$ is the true fluorophore distribution are shown in Table \ref{tab:errors}. 

\begin{table}[h]
\caption{Relative error ($\%$) in MAP estimates (\ref{mapref1}), (\ref{mapcem1}), (\ref{mapaem1}) for each 2D simulation test cases (test cases are numbered from top to bottom in Fig 1 and 2).} 
\label{tab:errors}
\begin{center}       
\begin{tabular}{|l|l|l|l|} %% this creates two columns
%% |l|l| to left justify each column entry
%% |c|c| to center each column entry
%% use of \rule[]{}{} below opens up each row
\hline
\rule[-1ex]{0pt}{3.5ex}  Case & $h_{\rm ref}$ & $h_{\rm cem}$ & $h_{\rm aem}$ \\
\hline\hline
\rule[-1ex]{0pt}{3.5ex}  1 & 42 & 42 & 59 \\
\hline
\rule[-1ex]{0pt}{3.5ex}  2 & 38 & 64 & 61 \\
\hline
\rule[-1ex]{0pt}{3.5ex}  3 & 44 & 100 & 66 \\
\hline
\rule[-1ex]{0pt}{3.5ex}  4 & 42 & 117 & 62 \\
\hline
\rule[-1ex]{0pt}{3.5ex}  5 & 43 & 116 & 66 \\
\hline
\end{tabular}
\end{center}
\end{table}

\subsection{3D simulations}

\begin{figure}[ht]
\centerline{\includegraphics[width=120mm]{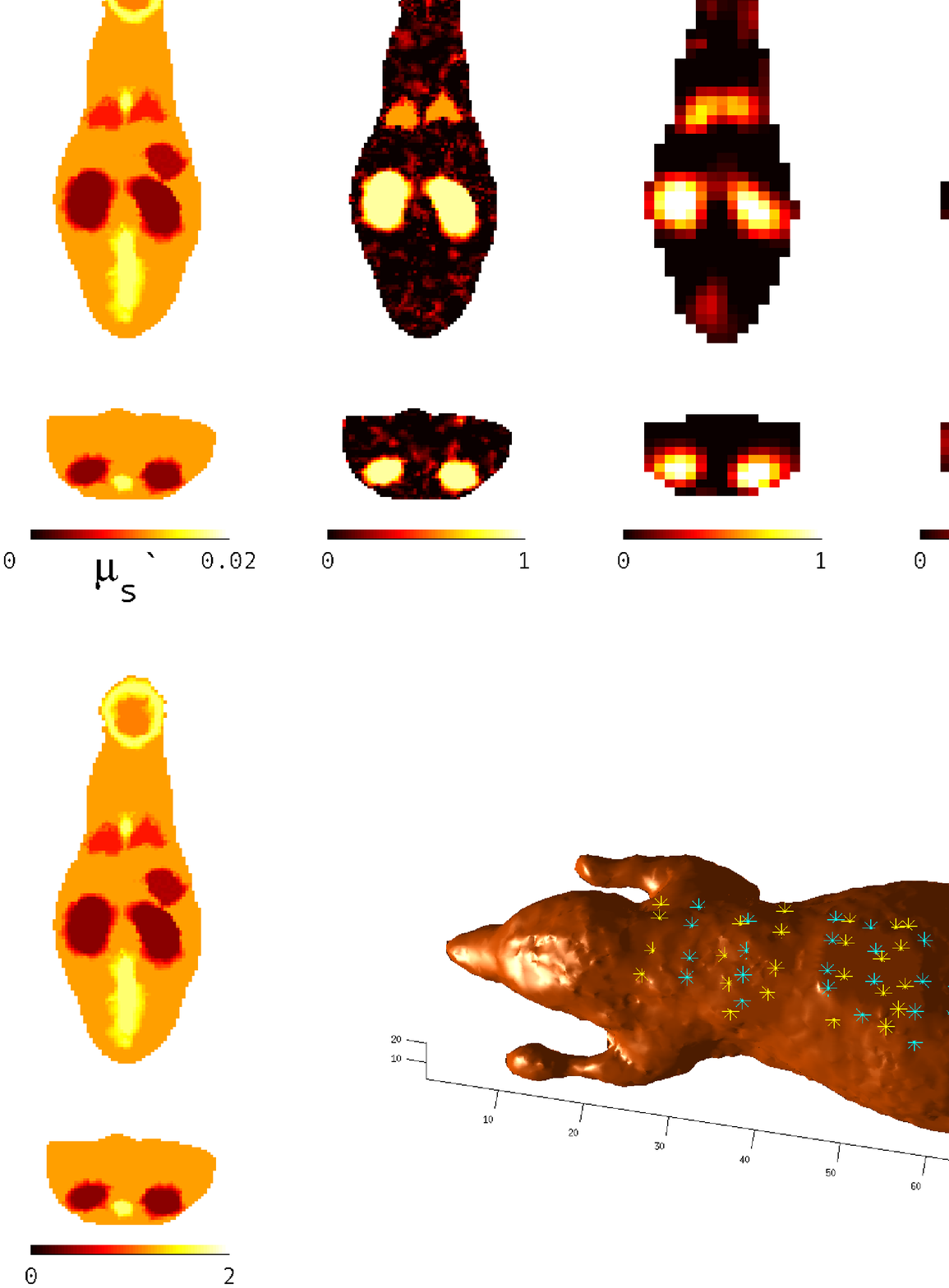}}
\caption{First column: target $\mua$ (top) and $\mus$ (bottom) at one horizontal and vertical slices of 3D mouse model. The second column shows the respective slices target fluorophore distribution $h$ of the body. Third column shows MAP-ref estimate using the correct absorption and scattering values (forward matrix $A(\mua,\mus)$). Fourth column shows MAP-CEM estimate using the incorrect absorption and scattering values (forward matrix $A(\muabar,\musbar)$). Fifth column shows MAP-AEM estimates using the same incorrect forward matrix $A(\muabar,\musbar)$. Bottom-right: Mouse surface with position of sources marked with yellow asterisk(*) and position of detectors marked with blue asterisks.}
\label{case2}
\end{figure}

The true absorption and scattering parameters $(\mua,\mus)$ from horizontal and vertical slices of target mouse model used in the simulations
are shown in Figure \ref{case2}, column 1: top (absorption) and bottom (scattering).
%These values are equivalent to the background values of true $(\mua,\mus)$.
The true target fluorophore distribution $h$ is shown in second column in Figure \ref{case2}. 
%As seen in the figure, the fluorophore concentration is specified in the lungs and kidneys of the mouse with finite contrast ($h$ $\textless$ 1 in this case). 
The nominal values $(\muabar, \musbar)$ that are used in the estimates $h_{\rm cem}$ and $h_{\rm aem}$  are homogeneous distributions with $\muabar(r) \equiv 0.01 \coefunit$ and $\musbar(r) \equiv 1 \coefunit$. The MAP estimates are:

\begin{description}
\item[(MAP-REF)] shown in column 3, Figure \ref{case2}. 
%The fixed values of the optical parameters 
%$(\mua, \mus)$ used in these estimates are the true target values of $(\mua, \mus)$. 
This corresponds to the reference estimate of conventional reconstruction when $(\mua,\mus)$ are known exactly.

\item[(MAP-CEM)] shown in column 4, Figure \ref{case2}. This corresponds to estimate with conventional reconstruction when the nominal values of $(\mua,\mus)$ are incorrect. 

\item[(MAP-AEM)] shown in column 5, Figure \ref{case2}. This corresponds to estimate with Bayesian approximation error 
model when the nominal values of $(\mua,\mus)$ are incorrect.

\end{description}

The results show that the Bayesian approximation error model efficiently compensates for the modeling errors caused by inaccurately known $(\mua,\mus)$ of the body and produce estimates that are qualitatively similar to the conventional estimates with exactly known $\mua$ and $\mus$. The inclusions in the fluorescent contrast are well localized with the Bayesian approximation error model and the estimates are relatively free of the artifacts that are present in the estimates $h_{\rm cem}$ with the conventional error model using the same incorrect values of $(\muabar,\musbar)$, 
see rows 2-5 in Figure \ref{case1} and columns 3-5 in Figure \ref{case2}.

A notable feature in the reconstructions is that estimate of $h$ has lower contrast in the reconstructions with the approximation error model than in the reconstructions with the conventional noise model. This can be explained by the fact that covariance of
the combined noise $e + \varepsilon$ is larger than the covariance of random noise (i.e.,
$\Gamma_e + \Gamma_\varepsilon > \Gamma_e$), implying that the relative weight
of the data residual term compared to the prior (or regularization) term becomes smaller in the MAP estimate with the approximation error model compared to the conventional noise model. Thus, the estimate gets, loosely speaking, drawn more strongly towards the prior mean, leading to a loss of contrast. This is also seen from the error estimates in Table \ref{tab:errors}. In the first row which corresponds to the case that $(\muabar,\musbar)$ are correct, the estimation error with the approximation error model is larger than with the conventional noise, and this discrepancy in the error arises from the 
lower contrast in the estimate of $h$ with the approximation error model.  
However, when the values $(\muabar,\musbar)$ are incorrect, rows 2-5 in Table \ref{tab:errors} and Figure \ref{case1},
the estimation error with the approximation error model is smaller than with the conventional noise model, and moreover, the  estimation error does not change much as the distance of $(\muabar,\musbar)$ from the true values $(\mua,\mus)$ increases when moving from row 2 to row 5 in the Table and Figure.

In Figure \ref{case2} we show simulation using a realistic 3D mouse model where we simulated high accumulation of fluorophores in the kidneys and lungs of the mouse. To make the simulation physiologically realistic we simulated background physiology by adding random fluorophore concentration $h = {\rm max}(h',0)$ where $h' \sim \mathcal{N} (0,\sigma^2\mathbb{I})$ with $\sigma = 0.2 \coefunit$, all over the mouse domain except the lungs and kidneys. The localisation of fluorophores in the kidney and lung positions can be seen in the MAP-ref estimates (column 3 Figure \ref{case2}). They appear slightly distorted in the MAP-CEM estimates (column 4 Figure \ref{case2}). However, the fluorophore concentrations are relatively better localised in the MAP-AEM estimate (column 5 Figure \ref{case2}) with a slight loss of contrast.

%
%This is presumably due to the increased likelihood variance ($\Gamma_e + \Gamma_\varepsilon > \Gamma_e$) in the %approximation error model compared to the conventional noise model.
%We also computed the reconstructions using the total variation prior model $\pitv$. An example of results, using the same data as in {\sc case 1} are shown in Figure \ref{case1tv}. These results indicate that sensitivity of the reconstruction to the modeling errors caused by the unknown $(\mua,\mus)$ is dependent on the prior model used in the reconstructions.

\section{Conclusions}
\label{sect:Discussion}

In this paper the recovery from errors caused by incorrectly modeled absorption and scattering in fDOT was considered. Born ratio is known to tolerate artefacts due to unknown absorption and scattering to some extent. However, in case the absorption and scattering properties are highly heterogeneous, incorrectly modelled absorption and scattering induces errors in the fDOT reconstructions. 

In this paper, the Bayesian approximation error approach was applied for the compensation of the errors caused by unknown absorption and scattering in fDOT. The modeling errors caused by the inaccurately known absorption and scattering were modeled as an additive modeling error noise in the observation model, and the posterior density model was then marginalized approximately over the unknown modeling errors by using a Gaussian approximation for the joint statistics of the primary unknown (fluorophore concentration) and the modeling errors. 

The approach was tested with 2D simulations with various target distributions of absorption and scattering. The results show that 
the approximation error model can efficiently compensate for the reconstruction artefacts caused by unknown absorption and scattering coefficients, even in the cases of highly heterogeneous absorption and scattering coefficients where the conventional 
estimates using the Born ratio contained severe artefacts. The approach was also tested with a 3D simulation using a mouse atlas. The MAP estimates using the Bayesian approximation error model show better localisation of the fluorophore concentration compared to the conventional estimate with the normalised Born approximation model. 

A tradeoff of the Bayesian approximation error model was found to be a small loss in contrast of the estimated fluorophore concentrations. We suggest that the approximation error model would be a feasible complement to the Born ratio model for handling the uncertainty related to the unknown absorption and scattering parameters in 
fDOT.
%fluorescence diffuse optical tomography.
  
\section*{Acknowledgements}

The work was supported by the Academy of Finland (projects 119270, 136220, 140984,
272803, and 250215 Finnish Centre of Excellence in Inverse Problems Research 2006-2011). The work was also supported by LASERLAB-EUROPE (grant agreement no. 284464, EC's Seventh Framework Programme).

\medskip
% The data information below will be filled by AIMS editorial staff
Received xxxx 20xx; revised xxxx 20xx.
\medskip

\end{document}